\documentclass[a4paper]{jpconf}
\usepackage{graphicx}
\usepackage{amsmath,amsthm,amsfonts}

\begin{document}
\title{Loop quantum cosmology of the Bianchi I model: complete quantization}
\author{M Mart\'in-Benito${}^{1}$, L J Garay${}^{2,3}$,  G A Mena Marug\'an${}^{2}$, E.~Wilson-Ewing${}^{4}$}

\address{
${}^{1}$ MPI f\"ur Gravitational Physics, Albert Einstein Institute,
Am M\"uhlenberg 1, D-14476 Potsdam, Germany.\\
${}^{2}$ Instituto de Estructura de la Materia, IEM-CSIC,
Serrano 121, 28006 Madrid, Spain.\\
${}^{3}$Departamento de F\'{i}sica Te\'{o}rica II, Universidad
Complutense de Madrid, 28040 Madrid, Spain.\\
${}^{4}$Centre de Physique Th\'{e}orique, CNRS-Luminy Case 907, 13288 Marseille Cedex 09, France.}

\ead{mercedes@aei.mpg.de, luis.garay@fis.ucm.es, mena@iem.cfmac.csic.es, wilson-ewing@cpt.univ-mrs.fr}

\begin{abstract}
We complete the canonical quantization of the vacuum Bianchi I model within the improved dynamics
scheme of loop quantum cosmology, characterizing the Hilbert structure of the physical
states and providing a complete set of observables acting on them. In order to achieve this task, 
it has been essential to determine the structure of the separable superselection sectors that arise
owing to the polymeric quantization, and to prove that the initial
value problem obtained when regarding the Hamiltonian constraint as an evolution equation,
interpreting the volume as the evolution parameter, is well-posed.\\

\end{abstract}

\section{Introduction}
Loop quantum cosmology (LQC) \cite{lqc1,lqc2,lqc3} adapts the techniques of loop quantum gravity
\cite{lqg} in the quantization of models with high degree of symmetry, such us homogeneous models.
Remarkably, the quantization of (homogeneous and isotropic) Friedman-Lema\^itre-Robertson-Walker
models within (the improved dynamics of) LQC succeeds in solving the singularity problem: the
classical big bang turns out to be replaced by a quantum bounce happening at Planck scales, and
no observable diverges in the quantum theory, as shown for the first time in the seminal
work of reference \cite{aps}.

In order to extent this quantization to (homogeneous but anisotropic) Bianchi models, we have
focused our attention in the simplest case: the vacuum Bianchi I model. Our interest in this
model also comes from the necessity of including inhomogeneities in LQC. Actually, with that aim 
we have analyzed from the LQC perspective the simplest inhomogeneous cosmologies: the linearly
polarized Gowdy model with three-torus topology \cite{hybrid3,hybrid4}. This model
can be regarded as a homogeneous Bianchi I background with three-torus topology filled with linearly
polarized gravitational waves traveling in a single direction.
The above quantization of the Gowdy model employs the
polymeric quantization of the Bianchi I model when representing the homogeneous background and,
therefore, first we need to have under control the Bianchi I model itself.

The quantization of the Bianchi I model in LQC is subject to several ambiguities.  One of them
concerns the representation of the curvature tensor of the connection in the quantum
theory. Different definitions of this object lead to different schemes of quantization, being the
most satisfactory one the {\sl improved dynamics} put forward in Ref.\
\cite{awe}, where although the Hamiltonian constraint was constructed, no further analysis of
the physical solutions was carried out. 

In this note we further study this quantization of the Bianchi I model. It summarizes some
results already presented in \cite{hybrid3,hybrid4}. We will start by adopting a different factor
ordering of that of \cite{awe} when symmetrizing the Hamiltonian constraint operator. With our
factor ordering the operator is well-defined in the octant of $\mathbb{R}^3$ given by the positive
eigenvalues of the operators representing the coefficients of the densitized triad.
In this way our model displays
sectors of superselection that are simpler than those of \cite{awe}. 
We will analyze the structure of the superselection sectors (aspect not studied in \cite{awe}) and
see that they are separable. 

The Hamiltonian constraint provides a difference equation in the volume. Then, it seems natural to
interpret it as an internal time and the constraint as an evolution equation with respect to it.
We will see that this interpretation is valid inasmuch as the
corresponding initial value problem is well-possed: physical solutions are completely determined by
a countable set of data given in the initial section of the volume (which displays a non-vanishing
minimum value). 
The previous result allows us to identify the physical Hilbert space with the Hilbert space of
initial data, whose
inner product is determined by imposing reality conditions in a (over) complete set of physical
observables. 

\section{Quantization of the model}

We consider the vacuum Bianchi I model with three-torus spatial topology. Then, we use global
coordinates $\lbrace t, \theta, \sigma, \delta \rbrace$, with $\theta,\sigma, \delta \in S^{1}$.
In order to prepare the model for its loop quantization, we describe it in the Ashtekar-Barbero
formalism \cite{lqg}. Using a diagonal gauge, the nontrivial
components of the densitized triad are $p_{i}/4\pi^{2}$, with $i=\theta, \sigma,
\delta$, and $c_{i}/2{\pi}$ are those of the $su(2)$ connection.
They satisfy $\left\{c_{i},p_{j} \right\}=8\pi G \gamma
\delta_{ij}$, where $\gamma$ is the Immirzi parameter and $G$ is the Newton constant. 
 The spacetime metric in these variables reads
 \begin{equation} \label{metric-bianchi}
 ds^2= -dt^2+\frac{|p_\theta p_\sigma p_\delta|}{4\pi^2}\left(
 \frac{d\theta^2}{p_\theta^2}+\frac{d\sigma^2}{p_\sigma^2}+\frac{d\delta^2}{p_\delta^2}\right).
 \end{equation}
The phase space is constrained by the Hamiltonian constraint
\begin{align}\label{eq:ligBianchi}
 C_\text{BI}=-\frac2{\gamma^2}\frac{c^\theta p_\theta c^\sigma p_\sigma+c^\theta p_\theta c^\delta
p_\delta+c^\sigma p_\sigma c^\delta p_\delta}{V}=0.
\end{align}
In this expression, $V=\sqrt{|p_\theta p_\sigma p_\delta|}$ is the volume of the compact spatial
sections.

In order to
represent the phase space in the quantum theory we choose the kinematical Hilbert space of the
Bianchi I
model constructed in LQC (see e.g. \cite{awe}), that we call $\mathcal{H}_{\text{kin}}$. We recall
that, on $\mathcal{H}_{\text{kin}}$, the operators $\hat{p}_{i}$ have a discrete
spectrum equal to the real line. The corresponding eigenstates,
$|p_\theta,p_\sigma,p_\delta\rangle$, form an orthonormal basis (in the
discrete norm) of $\mathcal{H}_{\text{kin}}$. Owing to this
discreteness, there is no well-defined operator representing the connection,
but rather its holonomies. They are computed along straight edges in the fiducial directions.
The so-called improved dynamics prescription
states that, when writing the curvature tensor in terms of holonomies, we have to evaluate them
along edges with a certain minimum dynamical (state dependent) length
$\bar\mu_i$. We use the specific improved dynamics
prescription put forward in \cite{awe}: the elementary operators which represent the matrix
elements of
the holonomies, called $\hat{\mathcal{N}}_{\bar{\mu}_{i}}$, produce all a constant
shift in volume $V$. In order
to simplify the analysis, it is convenient to relabel the basis states
in the form $|v,\lambda_\sigma,\lambda_\delta\rangle$, where $|v|$ is proportional to $V$
such
that the operators $\hat{\mathcal{N}}_{\pm\bar{\mu}_{i}}$ cause a shift on it equal to $\pm 1$,
and the parameters $\lambda_i$ are
all equally defined in terms of the corresponding parameters $p_i$, and verify
that $v=2\lambda_\theta\lambda_\sigma\lambda_\delta$.

Out of the basic operators $\hat{p}_i$, $\hat{\mathcal{N}}_{\bar{\mu}_{i}}$ we represent the
Hamiltonian constraint as an operator $\widehat{\mathcal C}$.
We choose a very suitable symmetric factor ordering such that $\widehat{\mathcal C}$ decouples the
states of $\mathcal{H}_{\text{kin}}$ with support in $v=0$, namely,
the states with vanishing volume \cite{hybrid3,hybrid4}. Moreover, our operator $\widehat{\mathcal
C}$ does
not relate states with different orientation of any of the eigenvalues of the operators
$\hat{p}_i$. Owing to this property we can then restrict the
domain of definition of $\widehat{\mathcal C}$ to e.g. the space spanned by the states
$|v>0,\lambda_\sigma>0,\lambda_\delta>0\rangle$. We call the resulting Hilbert space
$\mathcal{H}_{\text{kin}}^{+}$. Remarkably, for this restriction we do not need to impose any
particular
boundary condition or appeal to any parity symmetry. Note that both
$\mathcal{H}_{\text{kin}}$ and $\mathcal{H}_{\text{kin}}^{+}$ are
non-separable Hilbert spaces, feature not desirable for a physical theory. This problem is overcome
by the own action of the Hamiltonian constraint operator. Indeed, $\widehat{\mathcal C}$ defined on
$\mathcal{H}_{\text{kin}}^{+}$
turns out to leave invariant some separable Hilbert subspaces that provide superselection sectors.
In concrete, they are spanned by the states
$|v=\varepsilon+4n,
\lambda_\sigma=\lambda^{\star}_{\sigma}\omega_{\varepsilon},\lambda_\delta=\lambda^{\star}_{\delta}
\omega_ { \varepsilon } \rangle$. Here $\varepsilon\in(0,4]$ and $n\in\mathbb{N}$, and therefore
$v$ takes support in semilattices of constant step equal to $4$ starting in a minimum non-vanishing
value $\varepsilon$. In addition, $\lambda^{\star}_{a}$ ($a=\sigma,\delta$) is some positive real
number and
$\omega_{\varepsilon}$ runs over the subset of $\mathbb{R}^{+}$ given by
\begin{equation}
\mathcal
W_{\varepsilon}=\left\{\left(\frac{\varepsilon-2}{\varepsilon}\right)^{z}
\prod_{k}\left(\frac{\varepsilon+2m_k}{\varepsilon+2n_{k}} \right)^{p_{k}}\right\},\nonumber
\end{equation}
where $m_{k},n_{k},p_{k}\in \mathbb{N}$, and $z \in \mathbb{Z}$ when $\varepsilon>2$, while $z=0$
otherwise. One can check that in deed this set is dense in $\mathbb{R}^{+}$ and countable
\cite{hybrid3}. Therefore, any of these sectors provide separable Hilbert spaces
contained in $\mathcal{H}_{\text{kin}}^{+}$. We denote them as $\mathcal
H_{\varepsilon,\lambda_\sigma^\star,\lambda_\delta^\star}=\mathcal H_\varepsilon\otimes\mathcal
H_{\lambda_\sigma^\star}\otimes\mathcal H_{\lambda_\delta^\star}$.
Note that the
removal of the states with support in $v=0$ means that there is no analog of the cosmological
singularity (classically located in $p_i=0$) in our quantum theory. We thus solve the
singularity already at the level of superselection in a very simple way.

We then restrict the study to any of these superselection sectors, and expand a general solution in
the form
$
(\psi|=\sum_{v\in\mathcal
L_{\varepsilon}^+}\sum_{\omega_{\varepsilon}\in\mathcal
W_{\varepsilon}}\sum_{\bar\omega_{\varepsilon}\in\mathcal W_{\varepsilon}}
\psi(v,\omega_{\varepsilon}\lambda_\sigma^\star,\bar\omega_{\varepsilon}
\lambda_\delta^\star)
\big\langle
v,\omega_{\varepsilon}\lambda_\sigma^\star,\bar\omega_{\varepsilon}
\lambda_\delta^\star\big|
$.
One obtain that the constraint
$\big(\psi\big|\widehat{\mathcal C}{}^\dagger=0$ leads to a recurrence equation that relates the
combination of states
\begin{align}
\label{eq:combi}
\psi_+&(v+4,\lambda_\sigma,\lambda_\delta)=\psi\bigg(v+4,\lambda_\sigma,\frac{v+4}
{v+2}\lambda_\delta\bigg)+\psi\bigg(v+4,\frac{v+4}
{v+2}\lambda_\sigma,\lambda_\delta\bigg)+\psi\bigg(v+4,\frac{v+2}
{v}\lambda_\sigma,\lambda_\delta\bigg)\nonumber\\&+\psi\bigg(v+4,\lambda_\sigma,
\frac{v+2} {v}\lambda_\delta\bigg)+\psi\bigg(v+4,\frac{v+2}
{v}\lambda_\sigma,\frac{v+4} {v+2}\lambda_\delta\bigg)+\psi\bigg(v+4,\frac{v+4}
{v+2}\lambda_\sigma,\frac{v+2}{v}\lambda_\delta\bigg)
\end{align}
with data in the previous sections $v$ and $v-4$. Therefore, if
we regard $v$ as an internal time the constraint can be interpreted as an evolution equation in it.
It turns out that, when particularized to the initial time $v=\varepsilon$, the constraint simply
gives the combination of states $\psi_+(\varepsilon+4,\lambda_\sigma,\lambda_\delta)$ in terms of
some initial data $\psi(\varepsilon,\lambda'_\sigma,\lambda'_\delta)$.
From the combinations $\psi_+(\varepsilon+4,\lambda_\sigma,\lambda_\delta)$, it is possible to
determine any of the individual terms 
$\psi(\varepsilon+4,\lambda_\sigma,\lambda_\delta)$, since
the system of equations that relate the former ones with the latter ones is formally invertible. This issue
is very non-trivial. Actually, both the separability of the support of
$\lambda_a$ and the fact that it is dense in $\mathbb{R}^+$ have been essential to show it (the
details can be found in \cite{hybrid4}). Therefore, the initial value problem is well-posed and it
is indeed feasible to make the above interpretation of the constraint as an evolution equation in
$v$.

In conclusion, the physical solutions of the
Hamiltonian constraint are completely determined by the set of initial data
$
 \{\psi(\varepsilon,\lambda_\sigma,
\lambda_\delta)=\psi(\varepsilon,\omega_\varepsilon\lambda_\sigma^\star,
\bar\omega_\varepsilon\lambda_\delta^\star),\,
\omega_\varepsilon,\bar\omega_\varepsilon\in\mathcal W_{\varepsilon}\}                                      
$,
and we can identify solutions with this set. 
We can also characterize the physical Hilbert space as the Hilbert space of the initial data.
In order to endow the set of initial data with a Hilbert structure, one can take a complete set of
observables forming a closed algebra, and impose that the quantum counterpart of their
complex conjugation relations become adjointness relations between operators. This determines a
unique (up to unitary equivalence) inner product.

Before doing that, it is most convenient changing the notation. Let us
introduce the variables $x_a=\ln(\lambda_a)=\ln(\lambda_a^\star)+\rho_{\varepsilon}$. Note that
$\rho_{\varepsilon}$ takes values in a dense set of the real line, given by the logarithm of the
points in the set $\mathcal W_{\varepsilon}$. We will denote that set by $\mathcal Z_{\varepsilon}$.
Then, a set of observables acting on the initial data
$\tilde\psi(x_\sigma,x_\delta):=\psi(\varepsilon,x_\sigma,
x_\delta)$ is that formed by the operators
$\widehat{e^{ix_a}}$ and $\widehat{U}_a^{\rho_a}$, with $\rho_a\in\mathcal
Z_{\varepsilon}$ and $a=\sigma,\delta$, defined as
\begin{align}
 \widehat{e^{ix_\sigma}}\tilde\psi(x_\sigma,x_\delta)&={e^{ix_\sigma}}\tilde\psi(x_\sigma,x_\delta),
\qquad
\widehat{U}_\sigma^{\rho_\sigma}\tilde\psi(x_\sigma,x_\delta)=
\tilde\psi(x_\sigma+\rho_\sigma,x_\delta),
\end{align}
and similarly for $\widehat{e^{ix_\delta}}$ and $\widehat{U}_\delta^{\rho_\delta}$.
These operators provide an overcomplete set of observables and are unitary in
$\mathcal H_{\lambda_\sigma^\star}\otimes\mathcal H_{\lambda_\delta^\star}$, according with their
reality conditions.
Therefore, we conclude that this Hilbert space is precisely the
physical Hilbert space of the vacuum Bianchi I model.

\section{Conclusions}
Adopting the improved dynamics of reference \cite{awe}, we have completed the quantization of the
Bianchi I model in LQC providing the physical Hilbert space and a complete set of physical
observables, task so far not achieved. 

With that aim, first we have needed to determine the structure of the
superselection sectors that arise in the quantum model owing to the polymeric representation of the
geometry. These sectors display an involved structure concerning the
anisotropies, as a consequence of the complicated action of the Hamiltonian constraint on those
variables, inherent to the adopted improved dynamics. Actually, the values of the anisotropies run
over a dense set of the positive real line that turns out to be countable. In contrast, the volume
takes values in simple semilattices of constant step. As a result, the superselection sectors are
separable.

The Hamiltonian constraint provides a difference equation in the volume, and can be interpreted as
an evolution equation, being the volume the variable playing the role of the internal time.
The separability of the support of the anisotropies and the fact that it is dense in $\mathbb{R}^+$
are essential to show that the associated initial value problem is well-posed, and therefore
the above interpretation is valid. Since the initial data determines the solution, we can
characterize the physical Hilbert space as the Hilbert space of the initial data. The physical
inner product can be determined by imposing reality conditions on a complete set of observables. The
result is that the physical Hilbert space coincides with the tensor product of the superselection
sectors of the two anisotropy variables.

\ack This work was in part supported by the Spanish MICINN Projects No. FIS2008-06078-C03-03, No. FIS2011-30145-C03-02,  and
the Consolider-Ingenio Program CPAN No. CSD2007-00042.

\end{document}